\RequirePackage{fix-cm}
\documentclass[smallcondensed]{svjour3}     
\smartqed  
\usepackage{graphicx}
\usepackage{siunitx}

\begin{document}

\title{muCool: A novel low-energy muon beam for future precision experiments
}


\author{I. Belosevic     
		\and
        A. Antognini
        \and
        Y. Bao
        \and
        A. Eggenberger 
        \and
        M. Hildebrandt
        \and
        R. Iwai
        \and
        D. M. Kaplan
    	\and
        K. S. Khaw
        \and
        K. Kirch
        \and
        A. Knecht
        \and
        A. Papa
        \and
        C. Petitjean
        \and
        T. J. Phillips
        \and
        F. M. Piegsa
        \and
        N. Ritjoho
        \and
        A. Stoykov
        \and
        D. Taqqu
        \and
        G. Wichmann
}

%
\institute{I. Belosevic \at
              \email{ivanabe@phys.ethz.ch}           
           \and
           A. Antognini \and I. Belosevic \and A. Eggenberger \and R. Iwai \and K. S. Khaw \and K. Kirch \and F. M. Piegsa \and D. Taqqu \and G. Wichmann \at
              Institute for Particle Physics and Astrophysics, ETH Z\"urich, 8093 Z\"urich, Switzerland\label{label1}
              \and
              A. Antognini \and Y. Bao \and M. Hildebrandt \and K. Kirch \and A. Knecht \and A. Papa \and C. Petitjean \and N. Ritjoho \and
 A. Stoykov \at
             Paul Scherrer Institute, 5232 Villigen-PSI, Switzerland\label{addr2}
              \and
             D. M. Kaplan \and T. J. Phillips \at
             Illinois Institute of Technology, Chicago, IL 60616 USA\label{addr3}
              \and
             A. Papa \at
             Dipartimento di Fisica, Universit\`a di Pisa, and INFN sez. Pisa, Largo B. Pontecorvo 3, 56127 Pisa, Italy\label{addr6}
             \and K. S. Khaw \at
              \emph{Present address}: Department of Physics,University of Washington, Seattle, WA 98195, USA\label{addr5}
              \and F. M. Piegsa \at
              \emph{Present address}: Laboratory for High Energy Physics, Albert Einstein Center for Fundamental Physics, University of Bern, CH-3012 Bern, Switzerland\label{addr4}
              \and Y.Bao \at              
              \emph{Present address}: Institute of High Energy Physics, Chinese Academy of Sciences, Beijing 100049, China
}

\date{Received: date / Accepted: date}

\maketitle

\begin{abstract}

Experiments with muons ($\mu^{+}$) and muonium atoms ($\mu^{+}e^{-}$) offer several
promising possibilities for testing fundamental symmetries. Examples
of such experiments include search for muon electric dipole moment,
measurement of muon $g-2$ and experiments with muonium from laser spectroscopy to gravity experiments. These experiments
require high quality muon beams with small transverse size
and high intensity at low energy. 

At the Paul Scherrer Institute, Switzerland, we are developing a novel device that
reduces the phase space of a standard $\mu^{+}$ beam by a factor
of $10^{10}$ with $10^{-3}$ efficiency. The phase space compression
is achieved by stopping a standard $\mu^{+}$ beam in a cryogenic helium
gas. The stopped $\mu^{+}$ are manipulated into a small spot with
complex electric and magnetic fields in combination with gas density
gradients. From here, the muons are extracted into the vacuum and into a
field-free region. Various aspects of this compression scheme have
been demonstrated. In this article the current status will be reported.
\end{abstract}

\section{Introduction}
\label{intro}

Muon beams have important applications in particle and solid state
physics \cite{Nagamine2003a}. In solid state physics muon beams are used to study magnetic properties of materials using technique called muon spin rotation (${\rm \mu SR}$) \cite{Reotier1997}. In particle physics, experiments with
muons allow us to perform fundamental physics research, e.g. testing
electroweak interaction and searching for physics beyond the Standard
Model \cite{KlausP.Jungmann2002a}. Such experiments include measurements
of the anomalous muon magnetic moment \cite{Bennett2004a} or searches for the muon electric
dipole moment \cite{Bennett2009} and charged lepton flavor violation in forbidden muon
decays \cite{Adam2010,Berger2014}. 

Furthermore, the experiments with muonium atoms ($\mu^{+}e^{-}$ bound system)
open possibilities for high precision spectroscopy measurements \cite{Jungmann2006}.
The unique feature of the muonium atom is that hadronic (finite-size) effects are absent since it consists of only two leptons. This allows precise
tests of the bound-state QED and measurements of fundamental constants,
e.g. muon mass, muon magnetic moment and fine structure constant \cite{Cr}.
The muonium atom can also be used to search for CPT-invariance violation
and muonium-antimuonium conversion, which violates lepton number conservation.
The measurement of the gravitational acceleration of antimatter could
also be achievable with muonium \cite{Kirch2014,Kaplan2018}.

All these experiments could benefit from better quality (higher intensity, small phase space) muon beam at low energy (eV-MeV) \cite{KlausP.Jungmann2002a,Nagamine2014}. Furthermore, better quality muon beams would lead to better quality muonium beams.

At the Paul Scherrer Institute we are developing a device that transforms standard $\mu^{+}$ beam into a low-energy brilliant  $\mu^{+}$ beam \cite{Taqqu2006}. The working principle of this device is explained in the next section.



\section{Working principle of the proposed compression scheme}
\begin{figure*}
	\includegraphics[width=1.0\textwidth]{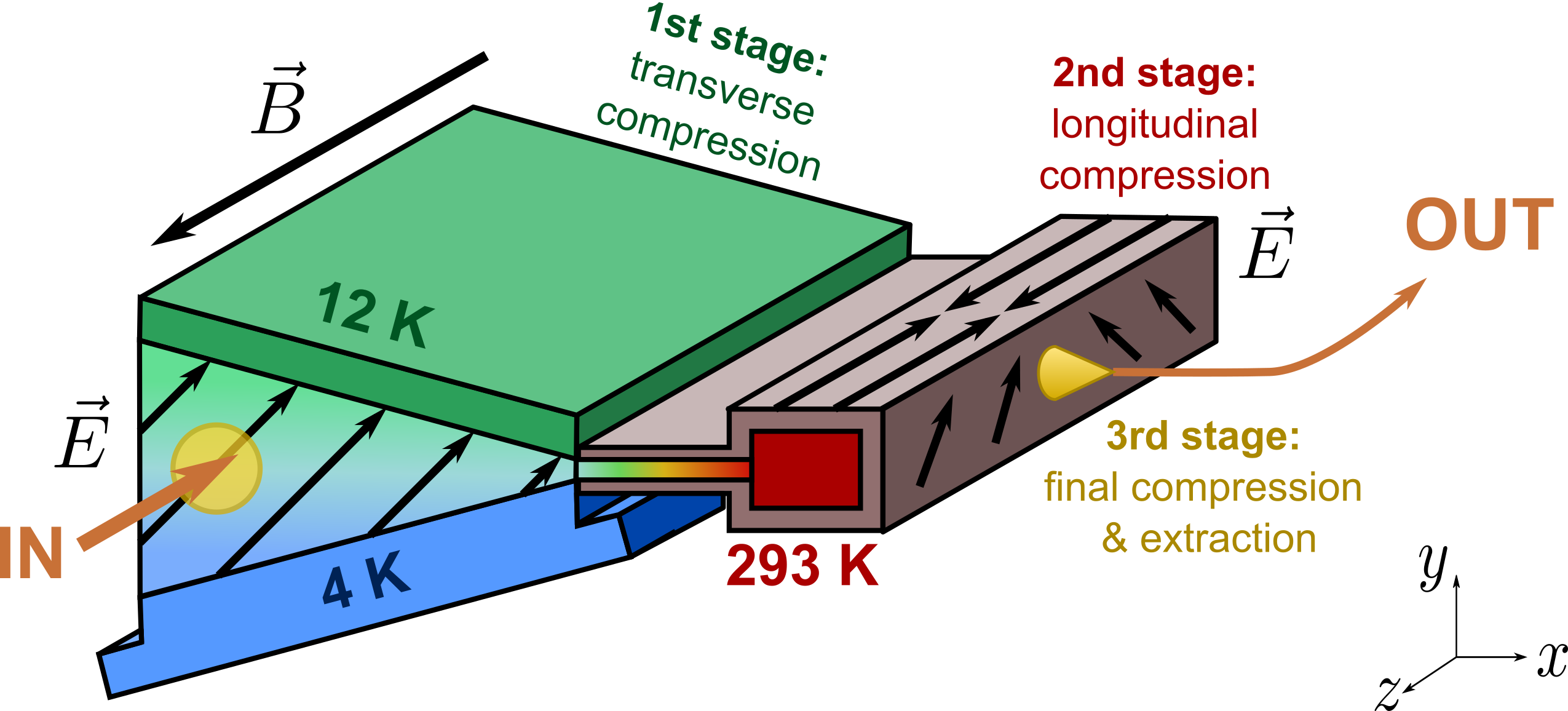}
	\caption{Scheme of the proposed muon compression beam line. Muons from the secondary
		$\mu^{+}$ beam enter the transverse compression stage, where they
		are first stopped in the helium gas and then compressed in transverse
		($y$) direction by using the combination of a vertical temperature
		gradient and the electric and magnetic fields. After that, they enter
		the longitudinal compression stage, where they are compressed in the
		longitudinal ($z$) direction and then extracted into the vacuum.}
	\label{fig:workingprinciple}       
\end{figure*}

In the proposed scheme a standard $\mu^{+}$ beam is stopped in a helium gas target consisting of three stages (see Fig. \ref{fig:workingprinciple}). Compression in all stages is achieved by manipulating stopped muons with complex electric and magnetic fields in combination with gas density gradients.

%
%

\subsection*{\textbf{1st stage: transverse compression}}
In the first stage, the standard $\mu^{+}$ beam
is stopped in a cryogenic helium gas target at few mbar pressure, placed in a magnetic field of $5$ T pointing in the $+z$-direction. An electric field $\vec{E}=(E_x,E_y,0)$ with $E_x=E_y \approx 1~\si{\kilo \volt \per \centi \meter}$  is applied perpendicular to the magnetic field.

In vacuum, applying such crossed electric and magnetic fields would cause muons to drift in $\vec{\hat{E}}\times\vec{\hat{B}}$ direction, performing cycloid motion characterized with the cyclotron frequency $\omega=eB/m$. However, the presence of the He gas leads to $\mu^{+}$-He collisions that  modify the muon motion. The deviation from the $\vec{\hat{E}}\times\vec{\hat{B}}$ direction (averaged over many collisions) will be proportional to the collision frequency $\nu_c$ between muons and He atoms, as described by the following equation \cite{Heylen1980}:

\begin{equation}
\tan\theta=\frac{\nu_{c}}{\omega},\label{eq:tan theta}
\end{equation}
where $\theta$ is angle of the muon drift velocity relative to the $\vec{\hat{E}}\times\vec{\hat{B}}$ direction.
Thus, we can manipulate the muon drift direction by changing the collision frequency $\nu_c$.

The collision frequency can be made position dependent by having different gas densities in different regions of our setup. In the transverse compression target this is achieved by keeping the upper wall of the target at $12$ K and the lower at $4$ K which creates a temperature gradient  and therefore also density gradient in the $y$-direction \cite{Wichmann2016a}. 

\begin{figure}
	




	\includegraphics[width=1.0\columnwidth]{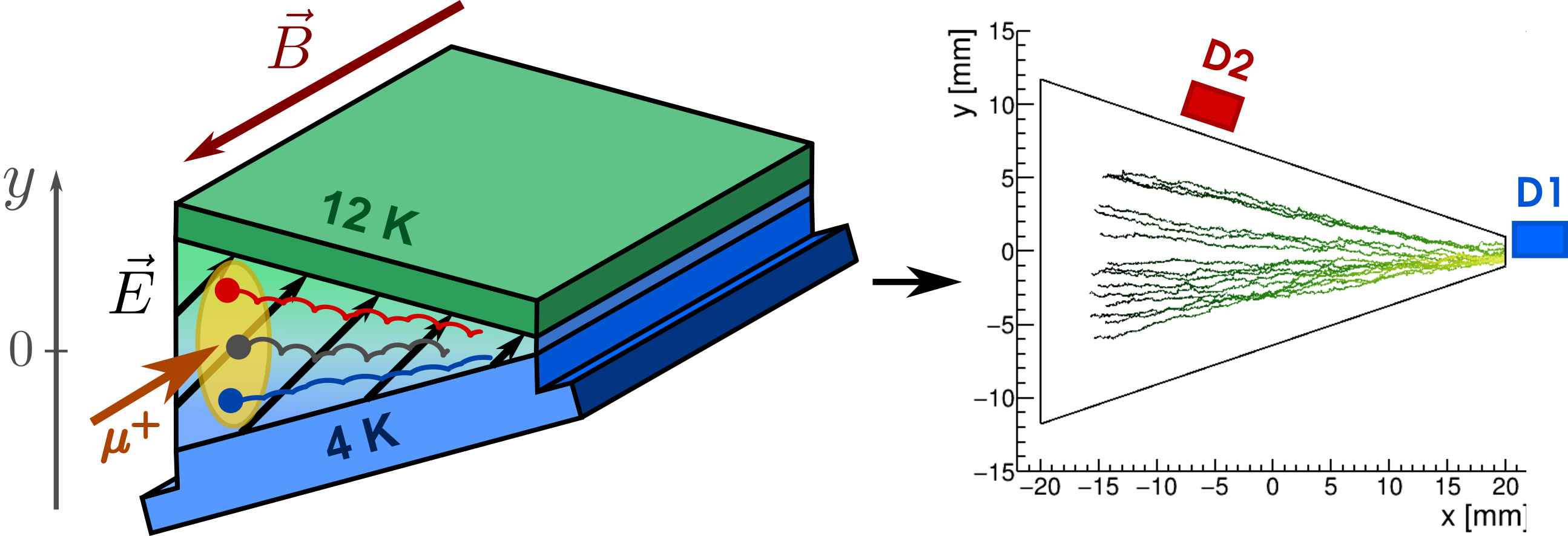}
	\caption{(Left) Sketch of the transverse target. Trajectories of muons for three different starting positions are sketched, as described in the text.
		(Right) GEANT4 simulation of the muon trajectories in the transverse target. Muons start around $x=-15$~mm and drift in $+x$-direction while simultaneously compressing in the $y$-direction.}
	\label{fig:trans_trajectories}       
\end{figure}

In the middle of the target (at $y=0$, see Fig.~\ref{fig:trans_trajectories} (left)), the gas density is chosen such that $\frac{\nu_{c}}{\omega}=1$.
According to the Eq.~(\ref{eq:tan theta}), at this condition the muons drift at \ang{45} angle with respect to $\vec{\hat{E}}\times\vec{\hat{B}}$ direction, which in our case corresponds to the $+x-$direction (see gray trajectory in Fig.~\ref{fig:trans_trajectories} (left)).

In the top part of the target, the gas density is lower, which means that $\frac{\nu_{c}}{\omega}<1$ and muons move essentially in the $\vec{\hat{E}}\times\vec{\hat{B}}$ direction. With our field configuration, this corresponds to muons moving $-y$-direction while drifting in the $+x$ direction (red trajectory in Fig.~\ref{fig:trans_trajectories} (left)).

In the lower part of the cell, at larger gas densities, $\frac{\nu_{c}}{\omega}>1$,  the muons drift mostly in electric field direction, i.e in $+y$ and $+x$ directions (blue trajectory in Fig.~\ref{fig:trans_trajectories} (left)). The result is transverse (in $y$-direction) compression of the muon beam. 

The GEANT4 simulation of the muon trajectories under such conditions is shown in Fig.~\ref{fig:trans_trajectories} (right). Muons start at around $x=-15$~mm with about 10~mm spread in the $y$-direction and drift in $+x$-direction while simultaneously compressing in the $y$-direction. At $x=20$~mm, the muon spread in $y$-direction is reduced to about 1~mm.

\subsection*{\textbf{2nd stage: longitudinal compression}}

\begin{figure}

	\includegraphics[width=1.0\columnwidth]{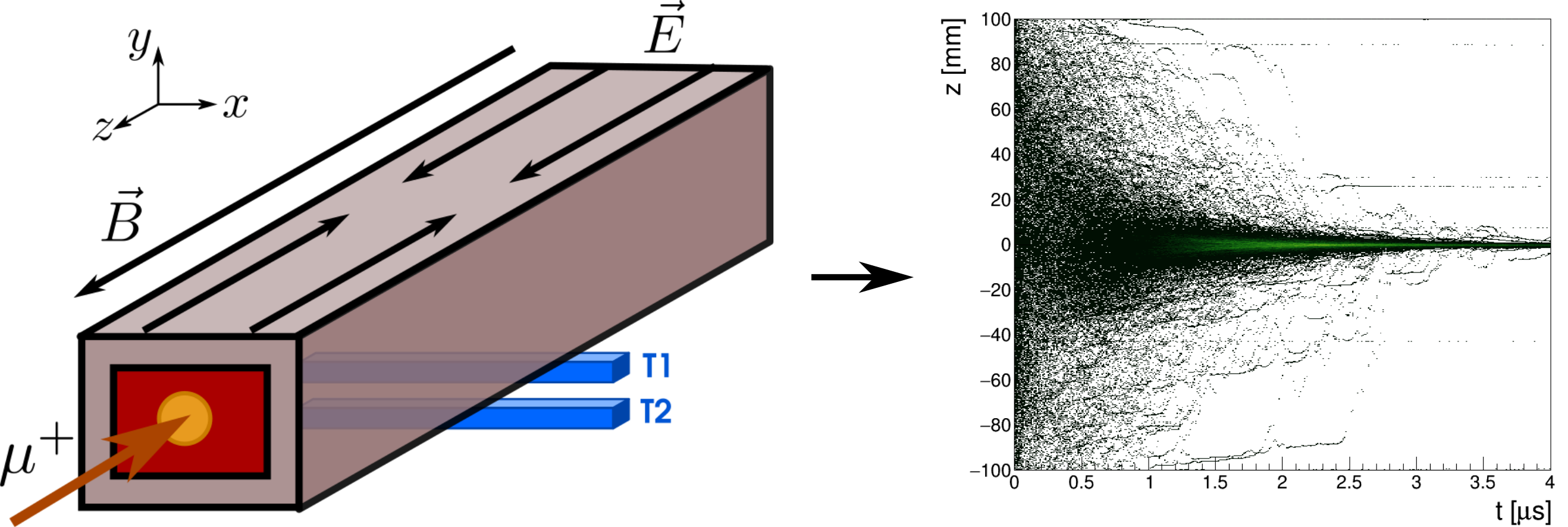}
	\caption{(Left) Sketch of the setup used to measure longitudinal compression.
		(Right) GEANT4 simulation of the muon $z$-position versus time in the longitudinal compression stage.}
	\label{fig:long_trajectories}       
\end{figure}

After the transverse compression stage, muons enter the second
compression stage, which is at room temperature. The electric field
now has a component parallel to the magnetic field and points
towards the center of the target, which causes a muon drift into the
center of the target, giving rise to the longitudinal (in $z$-direction) compression
of the muon beam (see Fig.~\ref{fig:long_trajectories}).

Additionally, there is a component of the electric field perpendicular to
the magnetic field, in $+y-$direction. Therefore, muons also drift
in $\vec{\hat{E}}\times\vec{\hat{B}}$ direction, which in this case points in
$+x-$direction, towards the final compression stage and extraction
into the vacuum.

Initially, the various stages of this compression scheme can be tested independently. Both longitudinal and transverse compression have been demonstrated, as presented in the sections~\ref{trans} and \ref{long}.

\section{Transverse compression demonstration}\label{trans}


To demonstrate transverse compression a 13~MeV/c muon beam was injected into the transverse target (as schematically shown in Fig.~\ref{fig:trans_trajectories} (left)).
The measurements were performed at the $\pi$E1 beam line at the Paul Scherrer Institute. 
This beam line provides about $2 \cdot 10^4~\mu^+$/s at 13~MeV/c.
Before entering the transverse target muons had to pass through a 55 \si{\micro \meter} thick entrance detector in front of the target, giving the initial time $t_0 = 0$. 

Several detectors were mounted around the target, as sketched in Fig.~\ref{fig:trans_trajectories} (right), to monitor the motion of the muons by detecting a positron from the muon decay. The detectors consisted of plastic scintillators in which a wavelength shifting fiber was glued. The wavelength shifting fibers transport the scintillating light from the cold to room temperature where the light is read-out by silicon photomultiplier (SiPM). Different detectors are sensitive to different regions of the transverse target (close to the detector). By recording the time difference between the positron hit in various detectors and the entrance detector,  time-spectra that carry the information about the muon motion are obtained.

\begin{figure}

	\includegraphics[width=1.0\columnwidth]{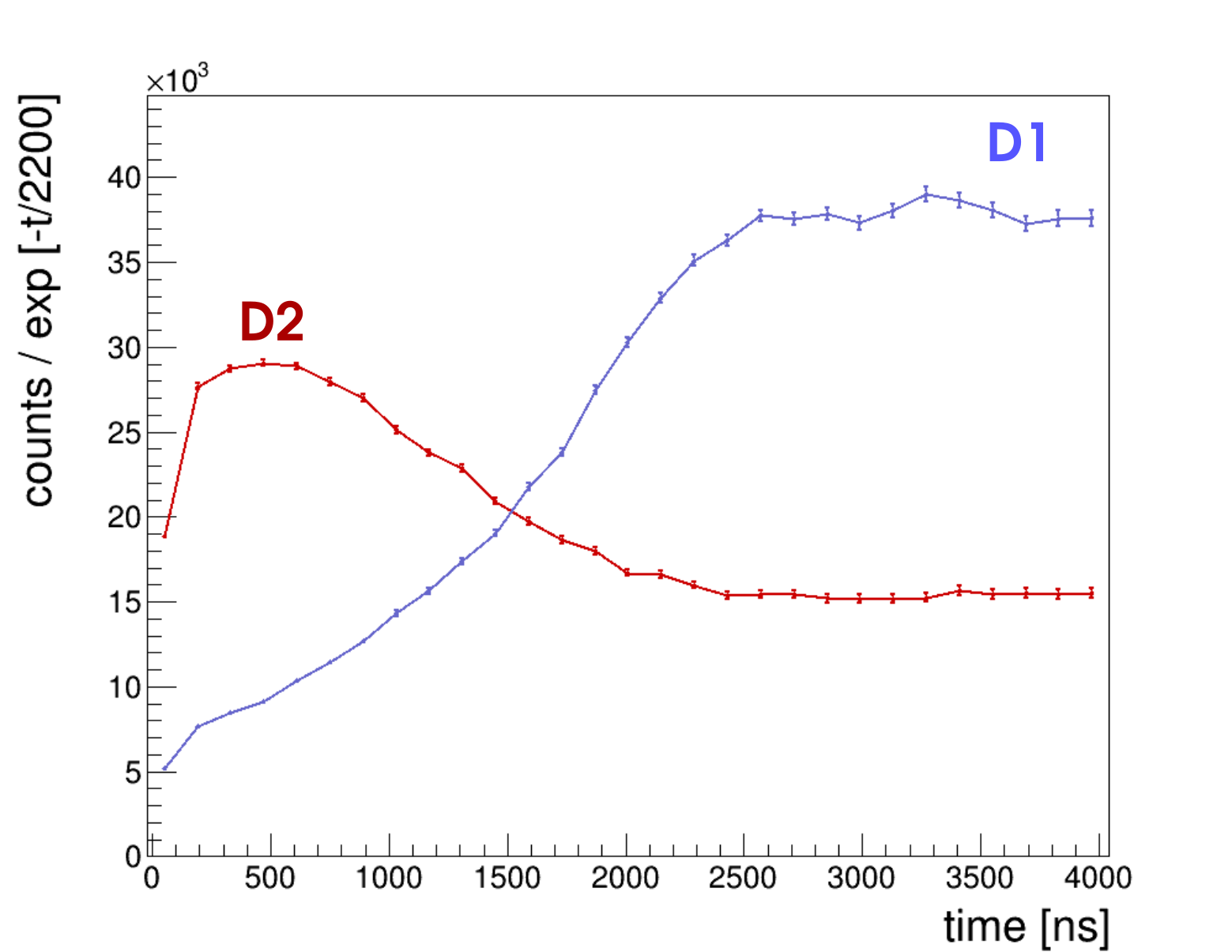}
	\caption{Measured time spectra of the detectors D1 and D2 for the transverse compression, using setup of Fig.~\ref{fig:trans_trajectories}. The lines connecting the points are drawn to guide the eye.
	}
	\label{fig:trans_data}       
\end{figure}

Two measured time-spectra are presented in Fig.~\ref{fig:trans_data} . The number of counts in each detector was corrected for the muon decay by multiplying the counts with ${\rm exp}(t/2198~\mathrm{ns})$. The number of counts in D1 (blue curve) increases with the time until about 2500~ns. This indicates that the muons were gradually moving closer to the detector acceptance region, i.e. closer to the tip of the target. After about 2500~ns the number of counts does not change with the time, which tells us that the muons were not moving anymore. On the other hand, the number of counts in the  D2 detector (red curve) first rises and then decreases with time, which tells us that the muons first entered and then exited the acceptance region of the detector (they "flew by" the detector).  These measured time-spectra are consistent with the simulated muon trajectories of Fig.~\ref{fig:trans_trajectories} (right).

\section{Longitudinal compression demonstration}\label{long}

Test of the longitudinal compression was done by  injecting a 11~MeV/c muon beam into a target containing 5~mbar of He gas at room temperature, as sketched in the Fig.~\ref{fig:long_trajectories} (left). The measurements were again performed at $\pi$E1 beam line at Paul Scherrer Institute. 
Similar as before, the initial time $t_0 = 0$ was produced by muon passing the 55 \si{\micro \meter} thick entrance detector in front of the target.

The longitudinal compression target with dimensions $12 \times 12 \times 200$~\si{\cubic \milli \meter} was lined with metallic electrodes that created a V-shaped electric potential with a minimum at the center of the target cell at $z=0$. Such potential caused the muons to move along the $\pm z$-direction towards the potential minimum, as shown in the Fig.~\ref{fig:long_trajectories} (right).

The muon motion was monitored with the two telescope detectors T1 \& T2 in coincidence that were placed at $z=0$, below the target, as shown in Fig.~\ref{fig:long_trajectories} (left). Massive brass shielding all around the target  ensured that coincidence hits in T1 \& T2  originated only from muons decaying within a small region close to the center of the target, between about $z=\pm3$~\si{\milli \meter}. From the time difference between the positron hit in T1 \& T2 in coincidence and the entrance detector, a time-spectrum is obtained. 

\begin{figure}

	\includegraphics[width=1.0\columnwidth]{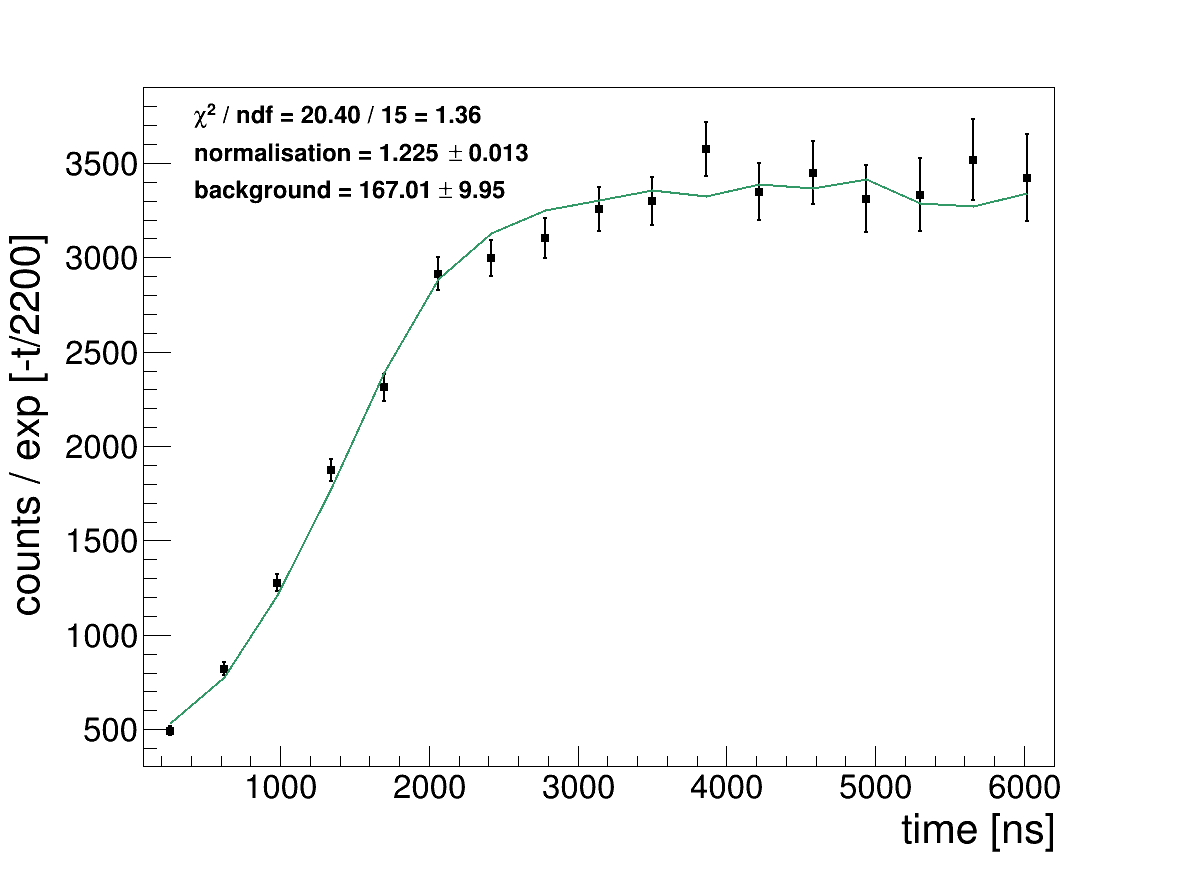}
	\caption{Measured time-spectrum for the longitudinal compression (points), using setup of Fig.~\ref{fig:long_trajectories}, and the corresponding simulated time-spectrum (green line).}
	\label{fig:long_data}       
\end{figure}

Figure ~\ref{fig:long_data} shows measured and simulated time-spectra when -500~V are applied in the center of the target cell. The measured number of counts (black points) increases with time which means that muons were moving closer to the acceptance region of T1 \& T2, corresponding to the center of the target. Thus, the 200~mm long muon swarm has been shrunk to below 6~mm length in $z$-direction within 4000~ns.  For more details see \cite{Bao2014,Antognini2018}.

The simulated time-spectrum (green line) was fitted to the data assuming two free parameters: normalization, to account for stopping and detection efficiency uncertainties, and flat background, to account for possible misalignment between the target and magnetic field axis. Measured time-spectra are in good agreement with the simulated time-spectra with reduced chi-square of 1.36 for 15 degrees of freedom. 

\section{Conclusions}

Several stages of the muon phase space compression beam line that we are developing at Paul Scherrer Institute have been tested. The muon transverse and longitudinal compression have been demonstrated independently. The measurements are consistent with GEANT4 simulations. The efficiencies assumed in the proposal~\cite{Taqqu2006} have been confirmed by measurements for both stages.
The merging of these two  stages and the extraction of the muon beam into vacuum are currently under development and will be tested in the near future.



\begin{acknowledgements}


	This work was supported by the SNF grants No. 200020\_159754 and 200020\_172639.

\end{acknowledgements}





\end{document}